\documentclass[prl,superscriptaddress,reprint,amssymb,aps,floatfix,showkeys]{revtex4-1}
\usepackage{amsmath}%
\usepackage{amsfonts}%
\usepackage{amssymb}%
\usepackage{graphicx}
\usepackage{color}
\usepackage{tabularx}
\usepackage{braket}

\renewcommand\Re{\operatorname{\mathfrak{Re}}}
\renewcommand\Im{\operatorname{\mathfrak{Im}}}

\begin{document}
\title{Jump Chaotic Behaviour of Ultra Low Loss Bulk Acoustic Wave Cavities}

\author{Maxim Goryachev}
\email{maxim.goryachev@uwa.edu.au}
\affiliation{ARC Centre of Excellence for Engineered Quantum Systems, University of Western Australia, 35 Stirling Highway, Crawley WA 6009, Australia}

\author{Warrick G. Farr}
\affiliation{ARC Centre of Excellence for Engineered Quantum Systems, University of Western Australia, 35 Stirling Highway, Crawley WA 6009, Australia}

\author{Serge Galliou}
\affiliation{Department of Time and Frequency, FEMTO-ST Institute, ENSMM, 26 Chemin de l'\'{E}pitaphe, 25000, Besan\c{c}on, France}

\author{Michael E. Tobar}
\affiliation{ARC Centre of Excellence for Engineered Quantum Systems, University of Western Australia, 35 Stirling Highway, Crawley WA 6009, Australia}

\begin{abstract}
We demonstrate a previously unobserved nonlinear phenomenon in an ultra-low loss quartz Bulk Acoustic Wave cavity ($Q>3\times10^9$), which only occurs below 20 milli-Kelvin in temperature and under relatively weak pumping. The phenomenon reveals the emergence of several stable equilibria (at least two foci and two nodes) and jumps between these quasi states at random times. The degree of this randomness as well as separations between levels can be controlled by the frequency of the incident carrier signal. It is demonstrated that the nature of the effect lays beyond the standard Duffing model. 

\end{abstract}
\date{\today}
\maketitle



Ultra low loss resonant systems are excellent tools to experimentally study nonlinear effects because of the higher probability of interaction between bounded energy quanta over a system nonlinearity. Common systems include photonic devices such as very high-Q Whispering Gallery Mode Resonators, which have drawn considerable attention in both the optical\cite{kerr,savchenkov2004,ilch2} and microwave\cite{Creedon:2012yg,PhysRevLett.108.093902} domains. Another experimental implementation of nonlinear low-loss systems encompass circuits based on Josephson Junctions, where a considerable degree of nonlinearity is apparent even near the quantum ground state and is the basis of circuit QED\cite{Blais:2004uq}.  
All these systems have potential applications in which the high degree of nonlinearity may be exploited, applications include cryptography\cite{Yang:1997qy}, nonlinear signal processing\cite{nonoptics}, quantum information processing\cite{QIP} as well as in study of fundamental principles of complex nonlinear systems\cite{nonapplic}. On the another hand, better understanding of these phenomena may also be used to avoid them when necessary.
Additionaly, low-loss phononic systems are capable of demonstrating some degree of nonlinearity. Although this is mostly limited to the well studied Duffing type nonlinearity\cite{duffing} arising from crystal anharmonicity. 
 Since nonlinear effects are more apparent in high Quality Factor systems, ultra low loss BAW devices at low temperature are good candidates for the study of mechanical nonlinearities beyond this model. These devices have the largest $Q\times f$ product among all the mechanical devices cooled to near their quantum ground state\cite{Kippen} exhibiting quality factors over 1 billion\cite{quartzPRL} at frequencies approaching $1$ GHz. As a result, BAW cavities at cryogenic temperatures demonstrate great potential for many physical applications\cite{ScRep}, in particular, as a mechanical system at the quantum limit\cite{Aspelmeyer:2008qc,schwab}. 

The Duffing nonlinearity plays a predominant role in quartz BAW devices. It is observed through the appearance of the hysteretic behaviour and third harmonic generation\cite{Tiers2,Tiers3,Goryachev:2zn,Nosek:1999cr}. Despite predictions of the period doubling bifurcation leading to chaotic behaviour\cite{Abe:1995xe,Novak:1982qq}, it has been never observed experimentally.
Nevertheless, recently some nonlinear effects beyond the Duffing model were discovered in Bulk Acoustic Wave (BAW) cavities at liquid helium temperatures\cite{quartzJAP}. The effect was explained by relatively high concentration of light impurities of the crystalline structure\cite{landaurumer1}. As a result, the effect could not be represented by the Duffing model whose main source is due to higher order terms in the crystal Hamiltonian. 
This work demonstrates another type of nonlinear behaviour discovered in a very pure cryogenically cooled ultra low loss BAW cavity, which only occurs in high-Q higher order overtones ($116$ and $135$ MHz) at relatively low drive powers below 20 mK.
 
 
The current experiment is made with a quartz SC-cut\cite{pz:1988zr} BVA-type\cite{1537081} plano-convex BAW Resonator. The plate device with a curved surface is $1$ mm thick, $30$ mm diameter electrode-separated disk cavities with higher grade surface polishing. This BAW resonator is initially designed to sustain slow shear vibration of $5$~MHz at room temperature, although its longitudinally-polorized modes exhibit extremely high values of quality factor at cryogenic temperatures\cite{quartzPRL,ScRep}. Such record high values of quality factor are achieved by the effective phonon trapping due to the curved plate surface. 

Plate BAW cavities could operate at various overtones (OT) of vibration with the quasi longitudinal (A-mode), quasi fast shear (B-mode) and quasi slow shear (C-mode) polarisations. The former type of mode exhibits significantly higher values of the $Q$ factor in phonon-trapping devices cooled to liquid helium temperatures and below. As a result, extremely high OTs (up to 227) could be excited and characterised\cite{quartzPRL}. 

The device is placed inside an Oxygen Free Copper block and cooled with the help of Dilution Refrigerator to approximately $17$mK. The resonator electrodes are connected to a single microwave transmission line split by several DC blocks. All the measurements are made in reflection by a Vector Network Analyser. Typically measurement results are presented in the form of $Z$-parameters, i.e. complex data characterising the device impedance at different frequencies near a resonance. 

Frequency stability of the measurements is controlled by a Hydrogen Maser. The maser guarantees fractional frequency stability no worse than $2\times 10^{-13}$ at 1 second and better than $2\times 10^{-15}$ at 1000 seconds of averaging times. For the OTs analysed here, the measurement setup provides frequency fluctuations on the order of 10 $\mu$Hz for averaging over one second. The influence of cryocooling system on instabilities of quartz BAW resonators have been investigated previously\cite{Goryachev:2012jx,Goryachev:2013ly}. 
 

Cryogenic BAW cavities are typically characterised by the Network Analysis method, which allows compensation for the connecting cable load. During this procedure, the incident probing signal is swept across the resonant frequency with the rate at which all transient effect could be neglected. Then a reflected signal is compared to the incident one at each frequency near the resonance. The correction procedure includes measurements of three calibration standards allowing the apparatus to remove the loading parasitic lines from the data. 
The measurements are made at the lowest accessible incident power of the order of $-45$~dBm in order to remove possible nonlinear behaviour\cite{Goryachev:2zn}. Although extremely high $Q$ values in excess of $1$ Billion may result in considerable degree of nonlinearity even at such low power. This is the case of the 37th and 39th OTs of the longitudinal mode of the device. These OTs at $f_A=116.16015112$ and $134.9927188$~MHz exhibit quality factors of $4\times 10^9$ and $3\times 10^9$ respectively\cite{quartzPRL} that are the highest values of $Q$-factors among all the modes at $17$~mK. Although the 39th OT exhibits very similar but less pronounced nonlinear behaviour, we demonstrate the results only for the 37th OT below. 

\begin{figure}[ht!]
     \begin{center}
            \includegraphics[width=0.45\textwidth]{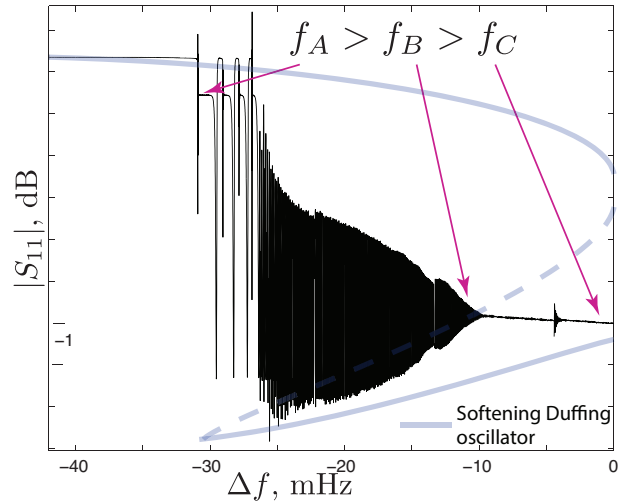}
            \end{center}
    \caption{Amplitude instabilities arising during characterisation of the 37th OT of the longitudinal mode in the quartz BAW cavity at $17$~mK with $-45$~dBm of the incident power. $\Delta f$ is the offset frequency from the $f_A=116.16015112$~MHz.}%
   \label{R001FR}
\end{figure}

During the frequency sweep in the vicinity of the 37th OT under the conditions described above, the amplitude instability behaviour is observed (Fig.~\ref{R001FR}). The sweep rate is kept as low as a few $\mu$Hz per second. As it is seen from the plot, the systems exhibits instabilities near the edge of the softening Duffing oscillator. Similar instabilities are observed on the phase plot. 

\begin{figure}[ht!]
     \begin{center}
            \includegraphics[width=0.5\textwidth]{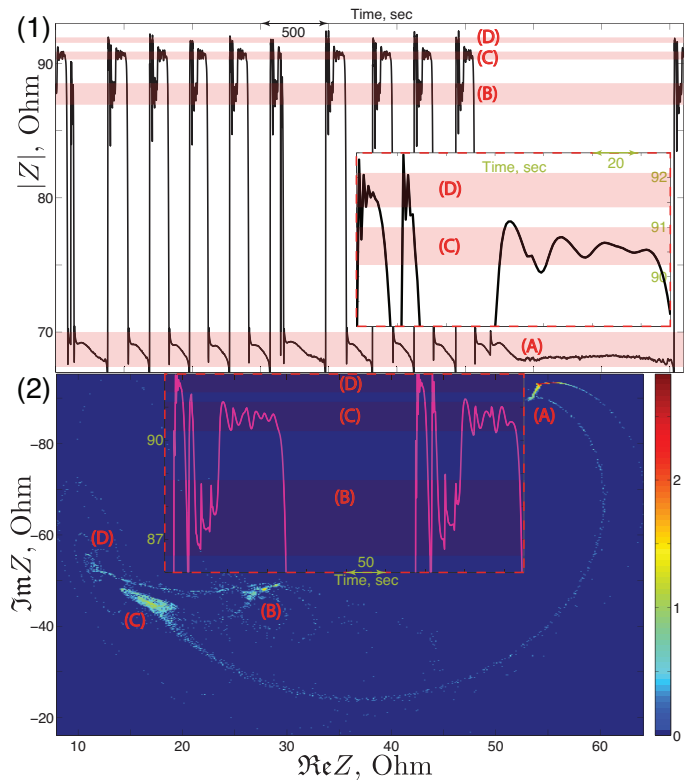}
            \end{center}
    \caption{(1) Time response of the magnitude of the scattering coefficient at pumping frequency $f_A$. (2) Two-dimensional histogram $\log_{10}N$ of the device impedance $(\Re Z, \Im Z)$. All insets show variation of $|Z|$ on different time scales. }%
   \label{R002FR}
\end{figure}

\begin{figure}[ht!]
     \begin{center}
            \includegraphics[width=0.5\textwidth]{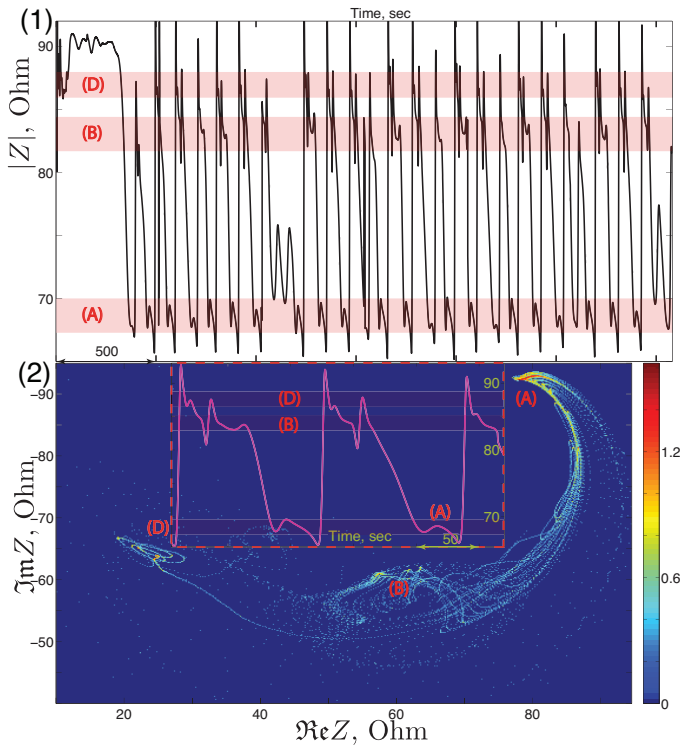}
            \end{center}
    \caption{(1) Time response of the magnitude of the scattering coefficient at pumping frequency $f_B$. (2) Two-dimensional histogram $\log_{10}N$ of the device impedance $(\Re Z, \Im Z)$. All insets show variation of $|Z|$ on different time scales.}%
   \label{R003FR}
\end{figure}

\begin{figure}[ht!]
     \begin{center}
            \includegraphics[width=0.5\textwidth]{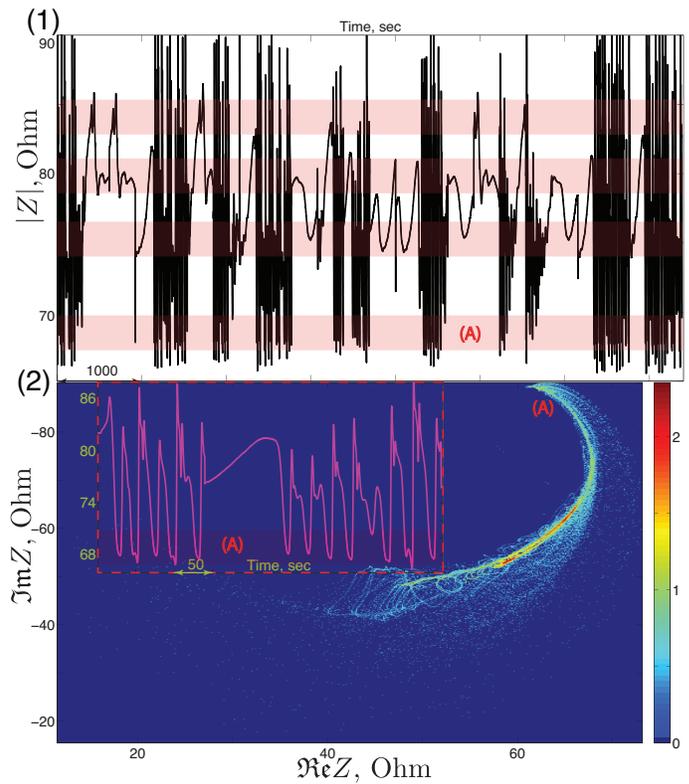}
            \end{center}
    \caption{(1) Time response of the magnitude of the scattering coefficient at pumping frequency $f_C$. (2) Two-dimensional histogram $\log_{10}N$ of the device impedance $(\Re Z, \Im Z)$. All insets show variation of $|Z|$ on different time scales.}%
   \label{R004FR}
\end{figure}

On the next step, the frequency of the incident signal is kept constant at some point of the instability region while the measurement setup was monitoring the both quadratures of the system response. The results are presented in two forms: time-series response for the magnitude of the device impedance $|Z|$ at a fixed frequency, a two-dimensional histogram of the result with respect to the real and imaginary part of the impedance in the logarithmic scale. The time series plots could be understood as demodulation of the resulting (current) signal with an original (voltage) signal related by the device impedance. In this case, $f_X$ plays a role of the carrier frequency. The histogram presents $\log_{10}N$ where $N$ is the number of measurement samples with the device impedance laying in the vicinity of $(\Re Z, \Im Z)$ point. These results for three frequencies $f_A$, $f_B$ and $f_C$ (see Fig.~\ref{R001FR}) are demonstrated in Fig.~\ref{R002FR}, \ref{R003FR}, \ref{R004FR} respectively.

In the ideal situation, when input signal frequency is fixed, one expects to observe the BAW device resonant impedance $Z$ independent of time after sufficiently long settling (transient) time. This situation is shown by a constant line on any of time-series plots (1) and a single bright spot on histograms (2) shown in Fig. ~\ref{R002FR}-\ref{R004FR}. Although, in our case due to the observed non-linear instability, $Z$ exhibits significant time fluctuations.  
Fig.~\ref{R002FR} reveals existence of at least four values (A)-(D) of the impedance where the resonant impedance $Z$ has higher chances to be observed. These quasi-levels are depicted as shadowed areas in (1) and appear as bright islands on the histogram (2). The brighter the island, the higher chances of an observer to measure this particular impedance at each given moment. Level (A) corresponds to the expected value of $|Z|$ at the resonance. Fig. ~\ref{R002FR} (2) shows that level (B) has an internal structure, three additional sub-levels.

Due to dynamic properties of the device, there are additional ringing (transient) effects leading to fluctuations of $Z$ around these four values as well as transitions between them. This results in a particular spiral-like appearance of the levels (D) and (C), which can be interpreted as stable foci. Contrary to that, level (A) and all sublevels of (B) behave as stable nodes of a nonlinear dynamical system. So, the whole behaviour could be characterised as jumping between quasi-levels of stable equilibria. 

The case of $f_A$ corresponds the the maximum separation between the stable states of the Duffing oscillator (Fig.~\ref{R001FR}) as well as the quasi-states in the current experiment (Fig.~\ref{R002FR}).
Increasing the offset frequency $\Delta f$ of the incident (carrier) signal, the gap between the two stable states of the Duffing oscillator and separations between the quasi-states of the actual mode decrease as a working point moves leftwards on Fig.~\ref{R001FR}. This can be observed also in Fig.~\ref{R003FR} (1) for the frequency $f_B$, where one of the quasi-states becomes indistinguishable from the others. As a result, the life time of the system in each quasi-state reduces leading to higher dispersion of the histogram (Fig.~\ref{R003FR} (2)). Consequently this leads to higher degree of randomness of the jumps between levels. These effects are amplified by further reduction of the carrier frequency. At frequency $f_C$ the system impedance demonstrate irregular motion with hardly distinguishable quasi-states (Fig.~\ref{R004FR} (1)). In fact the histogram Fig.~\ref{R004FR} (2) demonstrates very high dispersion of the quasi-states.

\begin{figure}[ht!]
     \begin{center}
            \includegraphics[width=0.5\textwidth]{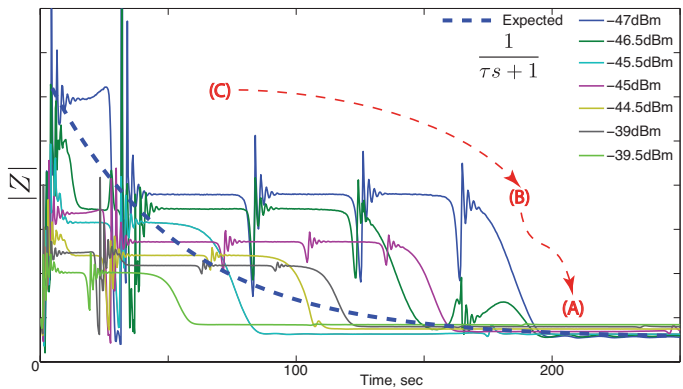}
            \end{center}
    \caption{Transient response of the mode impedance $Z$ for different values of the incident power near $f_A$. Input signal frequency is kept constant for all the measurements. }%
   \label{R005FR}
\end{figure}

Fig.~\ref{R002FR}-\ref{R004FR} demonstrate steady state behaviour of the system. In addition to that, one can consider a transient response of the system where the resonator exhibits a step-function change of the input carrier signal magnitude. Assuming that the carrier signal frequency equals to the effective resonance frequency, the system response in the magnitude-phase domain could be described by the first order filter response function $\frac{1}{\tau s+1}$, where $\tau = \frac{Q}{\pi f}\sim 150$ seconds is the resonance time constant and $s$ is the Laplace variable\cite{Goryachev:2011la,mybook}. The results of these measurements for different values of the incident power are shown in Fig.~\ref{R005FR}. Contrary to the expected exponential transient process with the time constant $\tau$, the system exhibits jumps between observed previously quasi-states at random times.  
The system demonstrates a decrease of the separations between the quasi-states with the increase of power. This could be related to the change of the frequency response of the corresponding Duffing oscillator with the applied power. 

The described effects are reproducible, they have been observed several times during separate cooldowns. The only requirement is operation at the coldest accessible temperature below $20$mK. Above this temperature only standard Duffing nonlinearity in the form of the hysteresis is observed. 


Although the behaviour of the ultra-high quality factor BAW cavity at mili-K temperatures under strong pumping is related to the dynamics of the Duffing oscillator, the system possesses some extra features beyond this model. Prior modelling of chaotic behaviour in this type of system\cite{Abe:1995xe,Novak:1982qq} predicts unobserved period-doubling phenomenon, but does not predict the quasi states observed here. Indeed, the nature of additional stable quasi-states and random jumps is so far not clear. In addition to this previously unobserved phenomenon, a successful theory has to explain the following peculiarities of the effect: 1) temperature threshold (the effect goes away for $T>20$~mK), 2) loss threshold (the effect goes away for the resonator with the degraded $Q$, as well as it is not observed for other lower-$Q$ modes), 3) upper power threshold. In particular we rule out thermal effects, such as those that occur in high-$Q$ optical resonators\cite{Creedon:2010fk,savchenkov2004,ilch2} at they do not result in appearance of extra quasi-states, upper power threshold and sharp temperature threshold. In contrast, thermal effects usually produces almost periodic (relaxation oscillation type) behaviour and become more pronounced at higher power. A certain degree of similarity can be observed between the described above  results, in particular, density plots in Fig.~\ref{R002FR}-{R004FR}, and Liouville/Husimi densities calculated for the driven quantum oscillator with chaotic parameters{\cite{quantchaos}}.

\begin{acknowledgments}
This work was supported by the Australian Research Council Grant No. CE11E0082 and FL0992016. Authors are thankful to Stefan Danilishin and Jean-Pierre Aubry for fruitful discussions. 
\end{acknowledgments}
\hspace{10pt}


%

\end{document}